\documentclass{PoS}

\usepackage{epsfig}

\usepackage{amssymb,amsmath,amsfonts,amsthm,mathrsfs}
\usepackage{array,multirow}

\usepackage{graphicx}

\graphicspath{{./figures/}}
\usepackage{fancybox}  
\usepackage{mediabb}
\usepackage{wrapfig}

\title{Development of Lattice QCD Tool Kit on Cell Broadband 
Engine Processor}

\ShortTitle{Lattice QCD on Cell/B.E.}

\author{\speaker{Shinji~Motoki}\\
Graduate School of Bio-Sphere Science, Hiroshima University,
Higashi-Hiroshima, Hiroshima, 739-8521, Japan   
        \\
        E-mail: \email{motoki-shinji@hiroshima-u.ac.jp}
}

\author{Yoshiyuki~Nakagawa
\\
Research Institute for Information Science and Education,
Hiroshima University, Higashi-Hiroshima, Hiroshima, 739-8521, Japan   
\\
E-mail: \email{nkgw@rcnp.osaka-u.ac.jp}
}

\author{Keitaro~Nagata
\\
Research Institute for Information Science and Education,
Hiroshima University, Higashi-Hiroshima, Hiroshima, 739-8521, Japan
\\
E-mail: \email{nagata@rcnp.osaka-u.ac.jp}
}

\author{Koichi~Hashimoto
\\
Fixstars Corporation,
Nisshin Bldg. 3F, 1-8-27, Kounan, Minato-ku, Tokyo 108-0075, Japan
\\
E-mail: \email{hashimoto@fixstars.com}
}

\author{Kiyoshi~Mizumaru
\\
Fixstars Corporation,
Nisshin Bldg. 3F, 1-8-27, Kounan, Minato-ku, Tokyo 108-0075, Japan
\\
E-mail: \email{maru@fixstars.com}
}

\author{Atsushi~Nakamura
\\
Research Institute for Information Science and Education,
Hiroshima University, Higashi-Hiroshima, Hiroshima, 739-8521, Japan
\\
E-mail: \email{nakamura@riise.hiroshima-u.ac.jp}
}

\abstract{
We report an implementation of a
code for SU(3) matrix multiplication on Cell/B.E.,
which is a part of our project, Lattice Tool Kit on Cell/B.E.. 
On QS20, the speed of the matrix multiplication on SPE
in single precision
is 227GFLOPS and it becomes 20GFLOPS\footnote{this vaule was remeasured and corrcted.} together with data transfer 
from main memory by DNA transfer, which is 4.6\% of the hardware peak speed
(460GFLOPS), and is 7.4\% of the theoretical peak speed of
this calculation (268.77GFLOPS).
We briefly describe our tuning procedure.
}

\FullConference{The XXVII International Symposium on Lattice Field Theory - LAT2009\\
		 July 26-31 2009\\
		 Peking University, Beijing, China}

\begin{document}

\section{Introduction}

This paper has  three objectives;
\begin{enumerate}
\item
We report a development of a SU(3) matrix multiplication 
code on the Cell/B.E.. This calculation is, needless to say, an essential
numerical part of any quench QCD calculation, and we 
expect that Cell/B.E. can be a good cost effective environment
for a quench QCD study such as the confinement, transport
coefficients etc..
\item
We report several techniques to  improve  our Cell program
performance, which we found during developing our code.
They are general and can be used to develop 
other scientific high-performance code.
\item
The experience reported here is a starting point to
develop a whole QCD simulation system on Cell/B.E.,
including the quark fermion matrix solver, 
which is now the most time
consuming part of lattice QCD.
Even a single Cell machine makes it possible for a researcher
or a small research group to calculate
quark dynamics.  And large scale Cell machine may work 
as a most powerful QCD machine\cite{Qpace09} 
\end{enumerate}

Lattice QCD society has been always looking for a powerful computer
for more than thirty years
to realize their dream, i.e., to simulate QCD on a computer,
to produce reliable data, and to understand non-perturbative
nature of QCD including the confinement, hadron interactions,
and quank-gluon plasma.  Machines for this aim were
VAX11, vector machines, parallel computers, clusters, GRID
and so on.

The Cell/B.E. has attracted much interest in scientific and 
engineering fields as a high performance machine
\cite{Qpace07,Motoki07,MILC,Qpace09}.
It has very high potential power, and is a much more reliable 
scientific computing system comparing current GP/GPU.

In order to extract its potential power,
programing on Cell/B.E. demands several cares, which were not
required on the traditional computers.This is because
Cell/B.E. is a machine of new architecture and has the following
features:
\begin{description}
\item[Multi-core computer:] A Cell/B.E. consists of
eight operation system processor cores 
called Synergistic Processor Element(SPE) 
and one system controll processor core 
called PowerPC Processor Element (PPE).
\item[SIMD operation:] The SPE is specialized to calculations 
in the use and has 
the new architecture with the ability 
of the SIMD operation.
\item[Small local memory LS:]
SPE has a local store (LS) of 256 KBytes
which worked as an inside memory of SPE.
\item[EIB connection and DMA:]
PPE and all SPE are connected by a high-speed bus 
called Element Interconnect Bus (EIB). 
The EIB is also connected to main memory 
and to an input and output device. Each 
processor core performs data access via the EIB.

SPE uses Direct Memory Access (DMA) transfer 
for data transmission. 
The DMA is used to forward data directly 
between memory and memory 
(or, memory and input/ output device). 

Although the EIB connection provides fast
transfer of the data,
the transfer time  should be hidden to get satisfactory performance. 

\item[Large number of registers:]
Each SPE has 128 general registers. It is important
to use this advantage to write an efficient
code.
\end{description}

In this report, 
we consider $SU(3)$ matrix multiplication,
$c^{(i)} = a^{(i)}\times b^{(i)}$ for 
$i = 0, N-1$. 
We show a result for $N=573,440$. 

The hardware peak speed of QS20, which has 2PPE and 16SPE,
is 460GFlops, while
theoretical peak speed of this calculation on Cell/B.E.  
is 268.769518291 GFLOPS.

\section{Tuning Procedure}

%
%
%
%
%
%
%
%
%
%
%

\subsection{Optimized Data Handling}

In order to extract SPE's calculational power,
the full use of its Single Instruction Multiple Data(SIMD)
function is essential. 
We must provide our input matrix data in a form which
fits the SIMD operation.
The SPE can handle several data at once by one instruction 
with vector data type. 
The data treated in  the Cell/B.E. are 16 bytes fixation, 
and it can handle four data at once by a single instruction 
in the single precision floating point arithmetic. 

When executing the DMA data transformation 
between the LS and main memory, 
a 128 byte aligned data structure is most efficient. 
The maximum amount of data at one transfer is 16 KByte, 
and it is desirable for data to fit 
this restriction. 

We pack the matrices, $a$ and $b$, of 16KByte in a structure. 
In order to fit the algorithm done on SPE, 
we separate the real and imaginary parts of a complex matrix, 
and pack 112 matrices. 

\begin{verbatim}
// DMA Send Data Struct(16kbyte Packed)
typedef struct _s_gprod0_send_t
{
    float          ar[3][3][112];
    float          ai[3][3][112];
    float          br[3][3][112];
    float          bi[3][3][112];
    unsigned char pading[256];
} s_gprod0_send_t;
\end{verbatim}

\subsection{
Effect of SIMDizaiton
}

The SIMD operation is the operation technique 
that can process plural data by one instruction.\cite{Ibrahim}
When we use only PPE, the above calculation takes 
365.080039 (msec), while it is 40.55844(msec) 
when we use SIMD (1SPE). 
Furthermore, when we employ 16SPE it takes 10.13458 (msec). 

\subsection{
Multi Buffering
}

In the parallel calculation, data
transmission often becomes the bottleneck, 
except trivially parallel applications.\cite{Williams,Kihara}
Therefore we must conceal the time for DMA transfer 
between the main memory and LS
\footnote{The calculation described here does not
require the data transfer among SPE's.
}.

For this purpose,
we adopt a technique called double-  or multi- buffering.
First let us consider a double buffering case
for transforming the matrices $a$ and $b$ from the main
memory to each LS of sixteen SPE's. 
We prepares two sets of buffers: 
\begin{itemize}
\item
We start to transfer the input data in the first buffer
by DMA to SPE's LS.
\item
When the data arrive, the SPE executes the matrix multiplication.
\item
Without interruption,
Memory Flow Controller(MFC) continues to send the next data in the second buffer.
This transfer time overlaps with the matrix-multiplication
calculation time. Thus part of the transfer time
is "hidden".
\item
MFC continues to send the next data in the first buffer.
\end{itemize}

\begin{figure}[ht]
\begin{center}
\resizebox{1.0\textwidth}{!} 
{\includegraphics{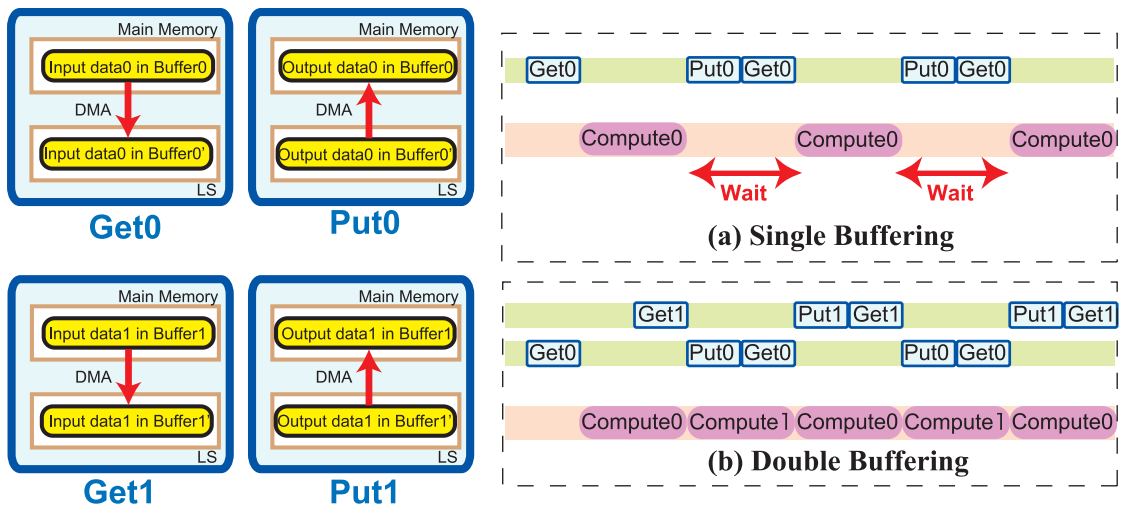}}
\vspace{-4.0cm}
\caption{Single and Double Buffering}
\label{Fig-DBuff}
\end{center}
\end{figure}

In Fig.\ref{Fig-DBuff}, 
we show a schematical diagram of (a)Single and (b)Double Buffering. 
When we use Single Buffering (16SPEs), it takes 10.1385 (msec),
while it takes 9.1241 (msec) when we employ Double Buffering(16SPEs). 
SIMD improves the efficiency three to four times. 
On the other hand, the effect of the double buffering is around 10$\%$.

\subsection{
Loop Unrolling and Software Pipelining
}

\begin{figure}
\begin{minipage}{0.47\hsize}\begin{center}
\resizebox{1.0\textwidth}{!}{\includegraphics
{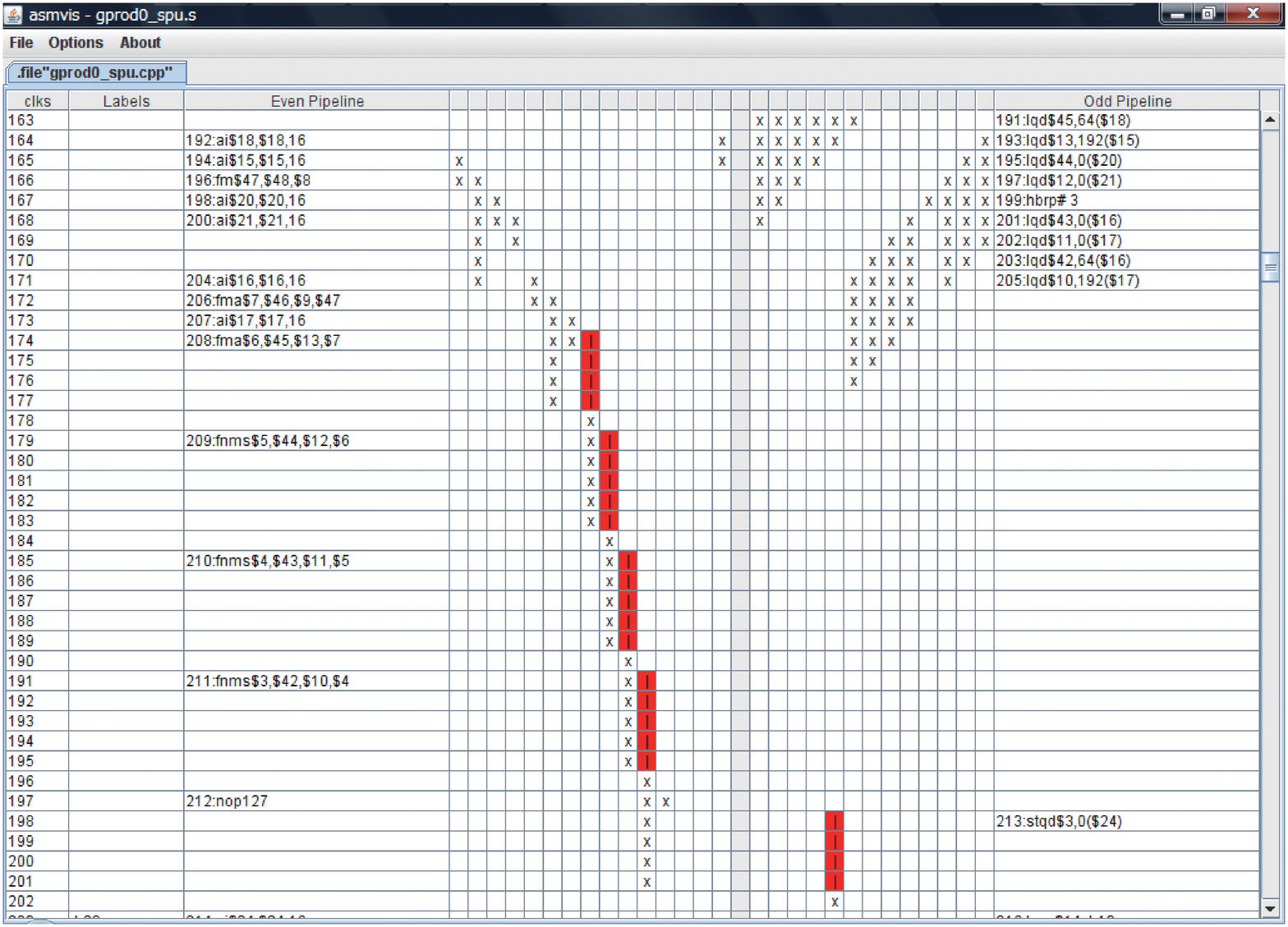}
}
\end{center}\end{minipage}
\hspace{0.03\hsize}
\begin{minipage}{0.47\hsize}\begin{center}
\resizebox{1.0\textwidth}{!}{\includegraphics
{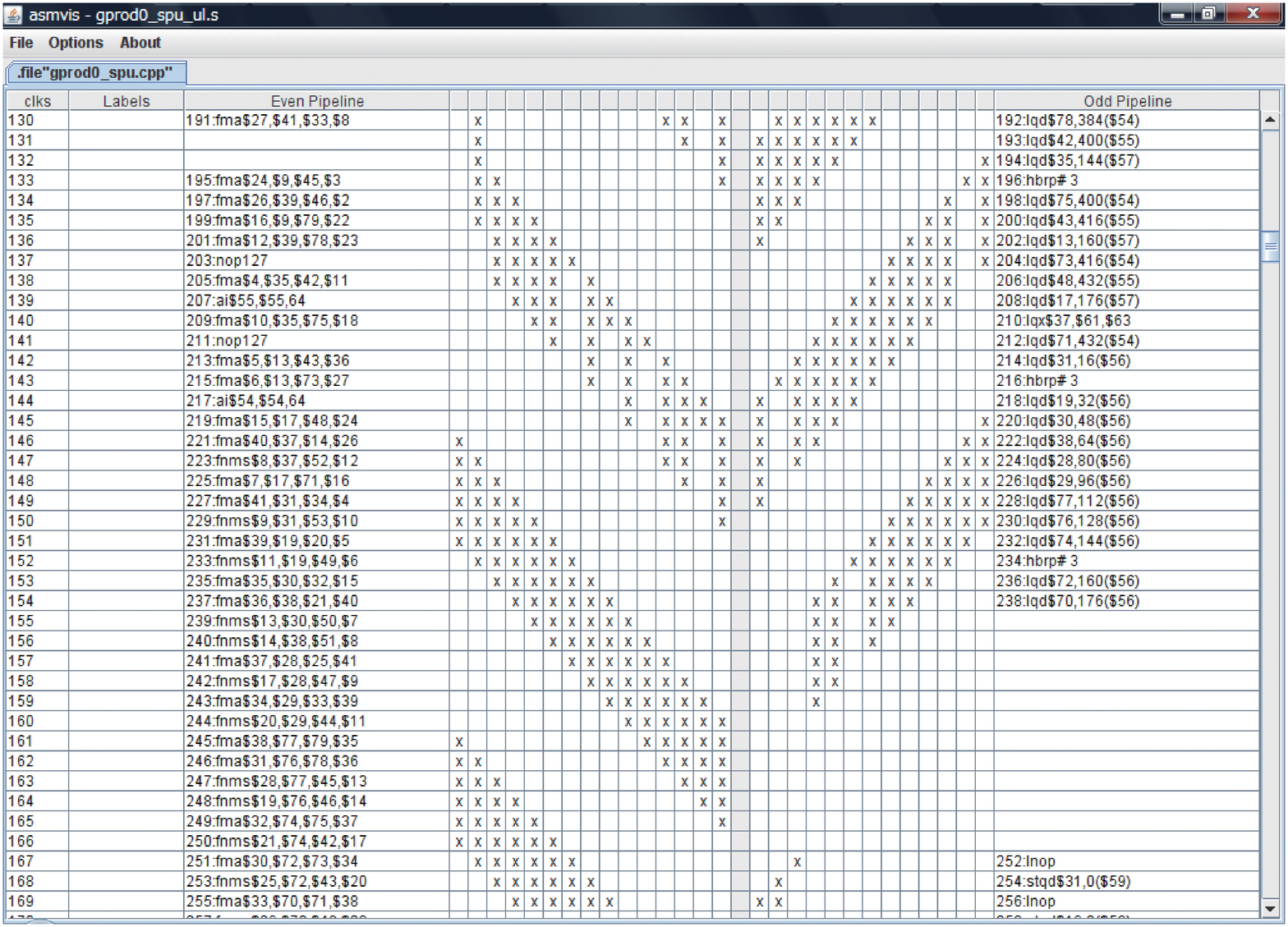}
}
\end{center}\end{minipage}
\vspace{+0.5cm}
\caption{
Dual-Pipeline optimization overview. \footnotesize{(left is Normal calculations, and right is dual-issue optimizaions)}
}
\label{Fig-Spu-timing}
\vspace{-0.5cm}
\end{figure}

The SPE has 128 general registers of 128 bit.
Thanks to the many registers,
loop unrolling results in a high-performance code 
(see Fig. \ref{Fig-Spu-timing}).
In addition, the SPE has the pipeline of 
two asymmetries, and two instructions 
can be executed at once. 
The speedup can be achieved by considering it. 

\vspace{2.0cm}
\begin{figure}[ht]
\begin{center}
\includegraphics[width=6cm,angle=-90]{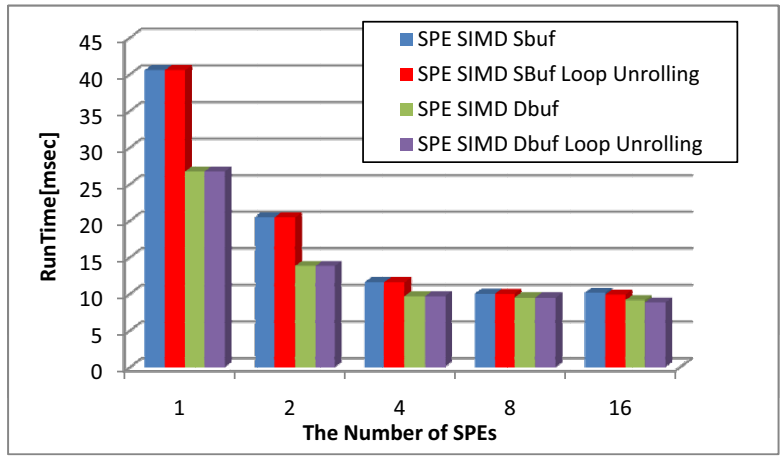}
\caption{Result of Loop Unrolling Calculations}
\label{Fig-LoopUnroll}
\end{center}
\end{figure}




Present compilers are not strong enough to find 
the most optimal level of the loop unrolling for 
a CPU of such many registers.
Thus, we must optimize a code by hand. 
We develop a loop by manual operation, and it widens 
the cord range where a compiler 
can optimize to use many registers. 
In addition, the dependency due to the register 
competition decreases, and one can conceal a stall. 

Furthermore the total road/store number in the loop decreases, 
and can conceal the access latency to LS. 
These are important advantages. 
We should avoid the resister competition. 
For this purpose we must keep it in mind that the load 
from LS to register needs six cycles. 
It makes a high performance code 
to use no variable which is just used, 
or to rearrange the order of operations.  

We show a calculation result of Loop Unrolling in Fig. \ref{Fig-LoopUnroll}. 
When we use 16 SPEs (SIMD and Double Buffering), 
it takes 8.8119 (msec) , This is equivalent to 13 GFlops.  
Therefore, we achieve about 41 times speedup 
in comparison with the speed in the PPE simple substance (365.080039msec, 
0.31GFlops) by tuning of the calculation technique in the SPE. 

\section{Discussions}

In this report, we briefly described our trials to run
the $SU(3)$ matrix multiplication code on Cell/B.E.
at high performance.
Combining several tuning techniques, we get over 20 GFLOPS(results of the latest our work),
which is about one tenth of the theoretical peak speed.

This is probably much better than expected in the community.
But we think more improvement is possible.
There are several points worth to be considered:
\begin{enumerate}
\item
Double buffering does not hide completely the transfer time.
Multi (more than four)
-buffering seems to be necessary.
\item
We waited the completion of the calculation, 
$c^{(i)} = a^{(i)}\times b^{(i)}$, and transfer back the result,
${c^{(i)}}$ to the main memory. 
This part has a room to be improved.
\item
There may be more efficient data structure for $a$,$b$ and $c$
for the best use of SIMD.
\end{enumerate}

In this experiment, we execute the calculation only once, and
therefore we create a thread once.  
In real calculations,
we must create threads many times, and it takes time to create
a thread.
Therefore, it is desirable to recycle the thread many times 
and to generate a thread once.

\section{Acknowledgement}
This work is supported by Grant-in-Aide for Scientific 
Research by Monbu-Kagaku-sho, Japan (20340055 ).

\end{document}